\documentstyle[prb,aps,12pt]{revtex}

\begin{document}
\title{On the effects of the magnetic field and the isotopic substitution upon
the infrared absorption of manganites}
\author{C.A. Perroni, G. De Filippis, V. Cataudella, G. Iadonisi, 
V. Marigliano Ramaglia, and F. Ventriglia}
\address{INFM  and Dipartimento di Scienze Fisiche, \\
Universit\`{a} degli Studi di Napoli ``Federico II'',\\
Complesso Universitario Monte Sant'Angelo,\\
Via Cintia, I-80126 Napoli, Italy}
\date{\today}
\maketitle

\begin{abstract}
Employing a variational approach that takes into account electron-phonon and 
magnetic interactions in $La_{1-x}A_xMnO_3$ perovskites with $0<x<0.5$, the 
effects of the magnetic field and the oxygen isotope substitution on the phase
diagram, the electron-phonon correlation function and the infrared absorption 
at $x=0.3$ are studied.
The lattice displacements show a strong correlation with the conductivity and 
the magnetic properties of the system.   
Then the conductivity spectra are characterized by a marked sensitivity to the 
external parameters near the phase boundary.

\end{abstract}


\newpage 
The perovskite oxides $La_{1-x}A_xMnO_3$ 
($A$ represents a divalent alkali element such as $Sr$ or $Ca$)
exhibit the colossal magnetoresistance ($CMR$) effect within the hole doping 
range $0.2 \leq x \leq 0.5$.\cite{jin,ramirez,dagotto}
These compounds are characterized by a 
transition from a metallic ferromagnetic ($FM$) low-temperature phase to an 
insulating paramagnetic ($PI$) high-temperature phase that is associated 
with dramatic changes in their electronic and magnetic properties.
 
These materials were first studied in the 1950's for their peculiar strong 
correlation between magnetization and resistivity. \cite{1} 
The ferromagnetic phase was explained by introducing the double exchange 
mechanism\cite{2,2bis} that takes into account the combined effect of the 
$Mn$ $e_g$ electron hopping between nearest neighbor sites and the very strong
Hund's exchange with the localized $Mn$ $t_{2g}$ electron spins.    
Later on, in addition to the double-exchange physics, a strong interaction  
between electrons and lattice distortions has been suggested in order to 
explain the $CMR$ phenomenon. \cite{3} 

At high temperatures, above the Curie temperature $T_c$, the Jahn-Teller 
polaron formation has been confirmed by many experimental measurements, 
\cite{salamon} in particular by the activated behavior of the conductivity, 
\cite{schiffer,worledge} the 
thermopower, \cite{jaim,palstra} the Hall mobility \cite{jaimi} and more 
directly by atomic pair distribution, \cite{billinge} electron paramagnetic 
resonance,\cite{shenge} x-ray and neutron scattering studies.
\cite{shimo,daii,teresa}
X-ray-absorption fine-structure ($XAFS$) spectra have found significant 
changes in the local structure of these compounds showing the crossover from 
large to small polarons across the metal-insulator ($MI$) transition and the 
direct relationship between lattice distortions, charge distribution 
and magnetism. \cite{tyson,booth1,booth,booth2,6}
Indeed the charge carriers partly retain their polaronic character even below 
$T_c$, as proved by neutron pair-distribution-function analysis 
\cite{louca}  and resistivity measurements. \cite{smoly}
Pseudogap features \cite{dessau1,dessau2} and conductivity spectra 
\cite{kim1,kim2,yoon,machida} have also been discussed in terms of a strong 
coupling to lattice distortions.
Finally the existence of the strong electron-phonon coupling and polarons has 
been demonstrated by the giant isotope shift of the Curie temperature.
\cite{5,isaac} Nowadays the key-role of the polaron formation in manganites 
is generally recognized.\cite{salamon,millis}

Recently, studies have stressed the intrinsic strong tendencies toward the 
phase separation in manganites.\cite{dagotto,yoon,zhou,9a,9b,9c}
Direct evidences for coexisting localized insulating and delocalized metallic 
components have been reported from tunnelling spectroscopy, neutron and 
electron measurements. 
\cite{daii,fath,adams,zuo}
The coexistence between hole-poor and hole-rich phases has been theoretically
studied by using exact numerical approaches on small lattices assuming 
classical Jahn-Teller phonons.\cite{dagotto,moreo} 
Within this approximation  the optical properties \cite{moreo2,millis2} have 
been studied but, at low temperatures, in the $FM$ phase, the 
narrow Drude peak, that is experimentally seen,\cite{kim1,kim2} cannot be 
obtained. 
   
In a recent paper based on a variational approach,\cite{cata} some of us have
shown that the quantum nature of the Jahn-Teller phonons 
and the polaron formation can be important to explain the experimentally 
observed tendency of manganites to form inhomogeneous magnetic structures near 
the phase boundaries. Employing the scheme proposed in that work, in a 
subsequent paper \cite{perroni} spectral and optical properties of manganites 
have been deduced at different temperatures for $x=0.3$ 
(around $x=0.3$ the $CMR$ effect is very pronounced in 
many manganites \cite{ramirez} ). 
By using the formalism of generalized Matsubara Green's functions
\cite{schna,loos1,loos2,kada,fehske}, the scattering rate of 
charges interacting with lattice and spin fluctuations has been 
determined and used to calculate the optical properties of the system.  
At low temperatures, in the $FM$ phase the system is characterized by two 
types of optical response: a Drude term and a broad absorption band due 
respectively to the coherent and incoherent motion of large polarons.
At high $T$ the infrared absorption is due to the incoherent small polaron 
dynamics. Upon cooling, the optical spectra have showed a transfer of spectral
weight from high to low energies filling up the low frequency optical gap 
present in the $PI$ phase in agreement with experimental data. 
\cite{kim1,kim2,okim}

In this paper, within the proposed variational approach, \cite{cata,perroni}
we calculate the lattice displacement probability distribution function 
($LDPDF$), a quantity measurable by $XAFS$ and introduced in order 
to study the lattice deformations due to the polaron formation.\cite{ranning}  
It allows to estimate the Debye-Waller ($DW$) factor for several $T$ 
at $x=0.3$. 
Near and within the $PI$ phase, the $LDPDF$ shows that sites occupied by an 
electron ($Mn^{3+}$) are characterized by strong lattice distortions, while 
sites without conduction electrons ($Mn^{4+}$) by a vanishing average static 
displacement.
Then we examine the effects of an external magnetic field on the phase diagram,
the lattice distortions, the infrared conductivity and the resistivity  
of the system at $x=0.3$. Near the phase boundary, the introduction 
of the field induces a transfer of spectral weight from high to low energies in
the conductivity spectra and, consequently, large variations in the 
resistivity.
In this approach, the $CMR$ effect is due to the subtle balance between two 
coexisting phases characterized by different lattice, spin and electronic 
properties. 
The comparison between the $DW$ factor in presence of the magnetic
field and the magnetoresistance ratio points out the correlation between 
lattice distortions, conductivity and magnetic properties.
Finally the isotopic substitution of the oxygen from $^{16} O$ to $^{18} O$ is
faced, finding that it increases the electron-phonon coupling reducing the 
Curie temperature $T_c$. In the vicinity of the $MI$ transition, the isotope 
effect causes large changes of the conductivity, so that it represents 
another example of the sensitivity of the system to the variation in external 
parameters. 

In Sec. I and II the variational approach and the lattice distortions are 
respectively discussed; 
in Sec. III the calculation of the damping of the particle motion and the 
optical properties are reviewed, \cite{perroni} 
in Sec. IV and V the introduction of an external magnetic field and the 
isotope effect are respectively examined.

\section{Variational approach}

We adopt a model that takes into account the double-exchange mechanism, the 
coupling of the $e_g$ electrons to lattice distortions, the super-exchange 
interaction between neighboring localized $t_{2g}$ electrons and the effects 
of an external magnetic field $\vec{h}_{ext}$. \cite{perroni,zhang}
The interaction to longitudinal optical phonons derives from the Jahn-Teller
effect that splits the $e_g$ double degeneracy.
Adopting the single orbital approximation (reasonable in the doping regime where $CMR$ occurs \cite{Yarla}),
the Hamiltonian reads
 
\begin{eqnarray}
H=&&-t\sum_{<i,j>} 
\left(\frac{S^{i,j}_0+1/2}{2 S+1}\right) c^{\dagger}_{i}c_{j}
 +\omega_0 \sum_{i}a^{\dagger}_{i}a_{i}
+g \omega_0 \sum_{i} c^{\dagger}_{i}c_{i} \left( a_{i}+a^{\dagger}_{i} 
\right)
  \nonumber \\
&& + \epsilon \sum_{<i,j>} \vec{S}_{i} \cdot \vec{S}_{j} 
-g_{s} \mu_{B}\sum_{i} \vec{h}_{ext} \cdot \vec{S}_{i}
- \mu \sum_{i} c^{\dagger}_{i} c_{i}.  \label{1r}
\end{eqnarray}
Here  $t$ is the transfer integral between nearest neighbor ($nn$)
sites $<i,j>$ for electrons occupying $e_g$ orbitals, 
$S$ is the spin of the $t_{2g}$ core states $\left( S= 3/2 \right)$,
$S^{i,j}_0$ is the total spin of the subsystem consisting of 
the spins localized on a pair of $ nn $ sites and the conduction electron, 
$c^{\dagger}_{i} \left( c_{i} \right)$ creates (destroys) an electron with 
spin parallel to the ionic spin at the i-th site. 
The first term of the Hamiltonian describes the double-exchange mechanism in 
the limit where the intra-atomic exchange integral $J$ is far larger than the 
transfer integral $t$. 
In the second term of eq.(\ref{1r}) $ a^{\dagger}_{i} \left( a_{i} \right)$ 
is the creation (annihilation) phonon operator at the site i,
$\omega_0=\sqrt{k/M}$ denotes the frequency of the optical phonon mode, 
with $k$ restoring force per length unit of the local oscillator and $M$ mass 
of an oxygen atom surrounding the manganese atom.
The dimensionless parameter $g$ indicates the strength of the
electron-phonon interaction within the Holstein model \cite{12}
\begin{eqnarray}
g=\frac {A}{\omega_0 \sqrt{2 M \omega_0} },
  \label{1r1}
\end{eqnarray}
where $A$ is the energy per displacement unit due to the coupling of the 
charge carriers with the lattice.
Finally in the Hamiltonian (\ref{1r}) $\epsilon$ represents the 
antiferromagnetic super-exchange coupling between two $nn$ $t_{2g}$ spins, 
$g_s$ the Lande's electron spin factor ($g_s \simeq 2 $), $\mu_B$ the Bohr 
magneton and $\mu$ is the chemical potential.

The hopping of electrons is supposed to take place between the equivalent 
$nn$ sites of a simple cubic lattice separated by the distance 
$|n-n^{\prime}|=a$. 
The units are such that the Planck constant $\hbar=1$, the Boltzmann constant
$k_B$=1 and the lattice parameter $a$=1.

We perform two successive canonical transformations to treat the 
electron-phonon interaction variationally. The first is the variational 
Lang-Firsov unitary transformation \cite{14,sil}
\begin{equation}
U_{1}=\exp \left[- g \sum_{j}  
\left( f c^{\dagger}_{j} c_{j} +\Delta \right) 
\left( a_{j}-a^{\dagger}_{j} \right) 
\right]                                            
\label{2r}
\end{equation}
where the parameter $f$  measures the degree of the polaronic effects and 
the parameter $\Delta$ denotes a displacement field
describing average static lattice distortions. The second
Bogoliubov-type transformation \cite{18} 
\begin{equation}
U_{2}=\exp \left[-\alpha \sum_{j} \left( a^{\dagger}_{j}
a^{\dagger}_{j} -a_{j}a_{j} \right)\right],  \label{3r}
\end{equation}
with $\alpha$ variational parameter, introduces correlations between the 
emission of successive virtual phonons by the electrons.

The transformed Hamiltonian $\tilde{H}=  U_{2} U_{1} H U_{1}^{-1} U_{2}^{-1} $ 
is difficult to treat exactly, so, in order to obtain a variational free 
energy, \cite{cata,perroni} we introduce a test Hamiltonian characterized by 
electron, phonon and spin degrees of freedom non mutually interacting

\begin{eqnarray}
H_{test} &=&
-t_{eff} \sum_{<i,j>} c^{\dagger}_{i}c_{j} +
\bar{\omega}_0 \sum_{i}a^{\dagger}_{i}a_{i}+
N \omega_0  \sinh^2\left( 2 \alpha\right)   
+N \omega_{0} g^{2} \Delta^{2}   \nonumber \\
&& 
-g_{s} \mu_{B}\sum_{i} \vec{h} \cdot \vec{S}_{i}
+\left( \eta -\mu \right) \sum_{i} c^{\dagger}_{i}c_{i}.  
\label{9r}
\end{eqnarray}
The quantity $t_{eff}$ denotes the effective transfer integral

\begin{equation}
t_{eff}= t \left\langle \left(\frac{S_0+1/2}{2S+1}\right)\right\rangle 
e^{-S_{T}}
\label{10r}
\end{equation}
where the symbol $<>$ indicates a thermal average and the quantity $S_{T}$ is

\begin{equation}
S_{T}=g^{2} f^{2} e^{-4 \alpha} \left( 2N_0+1 \right),
\end{equation}
with  $N_0$ the average number of phonons with frequency 
$ \bar{\omega}_0=\omega_0 \cosh(4 \alpha)$.
In the test Hamiltonian (\ref{9r}), $N$ is the number of lattice sites, $h$ 
the sum of the external magnetic field and the molecular magnetic 
field effective  in a cell containing two neighboring sites \cite{zhang1} and 
the quantity $\eta$

\begin{equation}
\eta = \omega_{0} g^{2} f \left( f-2 \right)+2 \omega_{0} g^{2} 
\left( f-1 \right) \Delta 
\label{8r}
\end{equation}
measures the electronic band shift due to the electron-phonon interaction.

We employ the Bogoliubov inequality in order to derive the variational free 
energy of the system
\begin{equation}
F \leq F_{test}+\langle \tilde{H}-H_{test} \rangle _{t}
\end{equation}
where $<>_t$ indicates a thermodynamic average made using the test 
Hamiltonian, so that the local spin dynamics is studied within a variational 
mean field theory.\cite{19}  The free energy per site becomes 
\begin{eqnarray}
\frac{F}{N}=&& 
f_{test}^{el} + T \log{\left(1-e^{-\beta \bar{\omega}_0}\right)}+ 
\omega_0  \sinh^2\left( 2 \alpha\right) 
+\omega_{0} g^{2} (1-f)^2 \rho^2 
- T\log{\nu_S}     \nonumber \\
&&
\pm \frac{\epsilon}{2} Z S^2 m^2_S +  T\lambda M 
\label{11r}
\end{eqnarray}
where the electron free energy $f_{test}^{el} $ is calculated considering the 
renormalized band $ \bar{\varepsilon}_{\bf{k}} = \varepsilon_{\bf{k}}+\eta $, 
where the band dispersion $\varepsilon_{\bf{k}}$ is  

\begin{equation}
\varepsilon_{\bf{k}}=-2 t_{eff}[cos(k_x)+ cos(k_y)+cos(k_z)].
\label{11ur}
\end{equation}
In eq.(\ref{11r}) $\beta$ is 
the inverse of the temperature, $\nu_S$ is the partition function of the 
localized spins, 
the top and bottom sign of $\epsilon$  hold, respectively, for the 
ferromagnetic and antiferromagnetic solutions of the localized spins and  
$Z$ indicates the number of nearest neighbors. The quantity  $\lambda$ is a 
dimensionless variational parameter proportional to the total magnetic field 
$h$
\begin{equation}
\lambda=\beta g_{s} \mu_{B} S h  
\label{11r1}
\end{equation}
and $M = B(\lambda)$ is the Brillouin function representing the 
ratio between the magnetization of the localized spins and the saturation 
magnetization.

In order to simplify the calculations, we consider a semicircular density 
of states

\begin{equation}
g(\epsilon)=\left( \frac{2}{\pi W^2} \right) 
\theta (W-|\epsilon| )   
\sqrt{ W^2-\epsilon^2 } 
\label{12r}
\end{equation}
where $W=Z t_{eff}$ is the renormalized band half-width and $\theta (x)$ is 
the Heaviside function.\cite{econo,georges} 

In the intermediate electron-phonon coupling regime, the free energy 
(\ref{11r}) gives rise to a region of coexisting phases 
characterized by different electron densities $\rho_1$ and $\rho_2$ 
and constituted by large and small polarons, $LP$ and $SP$, respectively.
Thus, near the metal-insulator transition, the system segregates 
in ferromagnetic and antiferromagnetic or paramagnetic domains of $LP$ and
$SP$.

The variational approach combined with the Maxwell construction allows to 
determine the fractions of volume of $LP$ and $SP$. Within the regime of the 
macroscopic phase separation ($PS$),
\cite{nagaev1,nagaev2} it is possible to calculate any thermodynamic property 
$B$ of the system by means of the linear combination of the properties 
$B_1$ and $B_2$ characteristic of the single homogeneous phases assuming the 
respective volume fractions  $\left( V_1/V \right)$ and $\left( V_2/V \right)$
as weights

\begin{equation}
B=\left( \frac {V_1}{V} \right) B_1+\left( \frac {V_2}{V} \right) B_2.
\label{14r}
\end{equation}
Therefore we assume that in this regime the properties of the system are 
independent of morphology of coexisting domains neglecting the surface energy 
cost. In the field of $CMR$ manganites, this two-fluid or two-component 
scenario has been proposed by other workers from a phenomenological point
of view. \cite{salamon,jaime}  
Furthermore, adopting this assumption, in a previous work \cite{perroni}  
the results for spectral properties and infrared absorption have turned out in
good agreement with experiments. 

In this paper the quantities are calculated taking into account
the phase diagram \cite{perroni} evaluated by using the following values of
the model parameters: $t=2 \omega_0$, $g = 2.8$ and $\epsilon= 0.01 t$.

\section{Lattice distortions}
In this section we deal with the lattice distortions in absence of the external
magnetic field. 

Within the above-mentioned variational approach, some of us have discussed the 
Jahn-Teller distortion of the $Mn^{3+}$ ion. \cite{cata}
It consists in an axial elongation of two $MnO$ bonds of the $MnO_6$ 
octahedra. \cite{6} 
Above a crossover temperature $T^* < T_c$, there are two distortions of this 
same type with a different degree of elongation along the axial direction, 
showing evidence of coexistence of large and small polarons.
The average displacement of the site $i$ from the equilibrium 
position, when one electron is on site $i$, is in agreement with the 
experimental data on the separations between local $Mn-O$ bond lengths. 
\cite{cata}     

Now we calculate the $LDPDF$ $P(X)$, that is more appropriate to make a 
comparison of the theory with the data obtained by $XAFS$ measurements. 
\cite{ranning} 
Considering the relevance of the axial elongation, we can confine ourselves 
to one spatial direction, so 

\begin{equation}
P(X)=<\delta (X-X_{i})>=\frac{1}{2 \pi} \int _{-\infty} ^{\infty}  
 dq   e^{i q X} <e^{-i q X_{i}}>,
\label{42r1}
\end{equation}
where $X_{i}=(a_{i}+a^{\dagger} _{i})/\sqrt{2 M \omega_0} $ is the 
displacement operator at the site $i$ along a reference axis. It has been 
evaluated performing the two canonical transformations (\ref{2r},\ref{3r}) and
making the decoupling \cite{perroni} in the electron and phonon terms through 
the introduction of $H_{test}$ (\ref{9r}). We get

\begin{equation}
P(X)=\frac {1}{\sqrt {\pi l}} 
\left[ \rho e^{-(x-b)^2 /l^2}+ (1-\rho)  e^{-(x-c)^2 /l^2} 
\right]
\label{42r2}
\end{equation}
where $\rho$ is the electron density, 
$ l=\sqrt { \frac { (2 N_0 +1) e^{4 \alpha} } {M \bar{\omega}_0 } }$,   
$b= g \sqrt { \frac {2} {M \omega_0 } } \left(  \Delta + f  \right) $ 
and 
$c= g \sqrt { \frac {2} {M \omega_0 } } \Delta $.
We obtain two terms: the first is due to the average occupation of the 
site $i$ and involves directly the polaronic distortion controlled by the 
variational parameter $f$, the second is due to the remaining configurations 
without electron on the site $i$ and it is related to the average lattice
displacement $\Delta = \rho (1-f)$. 
In Fig. 1(a) we report the distribution $P(X)$ for different temperatures at 
$x=0.3$. In the regime of coexisting phases $LDPDF$ takes contribution 
from both $LP-FM$ and $SP-PI$ distributions by means of the eq.(\ref{14r}). 
The $LP-FM$ phase is characterized by a single-peak $LDPDF$ centred about
the average lattice displacement $\Delta$ (here $f$ is smaller than $\Delta$). 
With increasing temperature, the spreading of $P(X)$ signals the PS regime 
and near $T_c$ two well-defined peaks appear giving a bimodal $LDPDF$. 
Near and in the insulating state, occupied sites ($Mn^{3+}$) have strong 
distortions while non-occupied sites ($Mn^{4+}$) are characterized by an 
average static displacement $\Delta$ equal to zero. \cite{tyson,booth}

We have derived the mean-square deviation of this distribution 
$\Delta x^2 = <x^2>-<x>^2$ that provides the $DW$ factor 
(see in Fig. 1(b) $DW$ as a function of $T$ at $x=0.3$). The $DW$ factor 
increases rapidly with temperature near $T_c$ in agreement with experimental 
data. \cite{booth1,booth2,dai}

\section{Damping and optical properties}

In this section we briefly review the calculations of the scattering rate and 
the infrared absorption due to lattice distortions and spin fluctuations. 
\cite{perroni} 

Retaining only the dominant autocorrelation terms at the second step of 
iteration \cite{schna,loos1,loos2,kada,fehske}, the self-energy
$\Sigma^{(2)} \left( {\bf k}, i \omega_n \right) $ and the scattering rate

\begin{equation}
\Gamma( {\bf k})=\tilde{\Gamma} ( {\bf k},\omega = \xi_{{\bf k}}  )
=-2 \Im \Sigma^{(2)}_{ret} 
\left( {\bf k}, \omega= \xi_{{\bf k}}  \right)
\label{42r}
\end{equation}
are derived.
Making the expansion into the series of multiphonon processes, 
the scattering rate is expressed as

\begin{equation}
\Gamma( {\bf k})=\Gamma(\xi_{{\bf k}})= 
\Gamma_{1-phon}(\xi_{{\bf k}})+ \Gamma_{multi-phon}(\xi_{{\bf k}}) + 
\Gamma_{Spin-Fluct}(\xi_{{\bf k}})
\label{44r}
\end{equation}
where $\Gamma_{1-phon}$ is the contribution due to single phonon
processes, $\Gamma_{multi-phon}$ represents the scattering rate by 
multiphonon processes and $\Gamma_{Spin-Fluct}$ denotes the damping term by 
spin fluctuations. At low $T$, in the $LP$ phase, single phonon 
emission and absorption represent the main mechanism of scattering. 
Furthermore, at higher temperatures the damping due to spin fluctuations is
effective in the energy range around the chemical potential $\mu$.

The scattering rate turns out fundamental to derive the infrared absorption of
the system. The real part  of the conductivity is obtained by the 
current-current correlation function 

\begin{equation}
\Re \sigma_{\alpha,\alpha}(\omega)= 
- \frac{ \Im \Pi_{\alpha,\alpha}^{ret}(\omega) }{\omega}
\label{54r}
\end{equation}
where the current operator $j_{\alpha}$ suitable for the Hamiltonian (\ref{1r})
is

\begin{equation}
j_{\alpha}=ite \sum_{i,\delta}  \delta_{\hat{\alpha}}
\left( \frac{S_0^{i+\delta \hat{\alpha},i}+1/2}{2 S+1} \right)
c^{\dagger}_{i+\delta \hat{\alpha}} c_i.
\label{56r}
\end{equation}
The conductivity can be expressed as sum of two terms 

\begin{equation}
\Re \sigma_{\alpha,\alpha}(\omega)= 
- \frac{ \Im \left[ \Pi_{\alpha,\alpha}^{ret (1)}(\omega)+ 
\Pi_{\alpha,\alpha}^{ret (2)}(\omega) \right]} {\omega}
=\Re \sigma^{(band)}_{\alpha,\alpha}(\omega)
+ \Re \sigma^{(incoh)}_{\alpha,\alpha}(\omega).
\label{68r}
\end{equation}
The first term $\Re \sigma^{(band)}_{\alpha,\alpha}$ represents the 
band conductivity because the absorption is not accompanied by processes 
changing the number of phonons. 
On the other hand, the second term marked by the apex ``incoherent'' 
in eq. (\ref{68r}) derives from inelastic scattering processes of emission 
and absorption of phonons.
The band conductivity is derived as 

\begin{equation}
\Re \sigma^{(band)}_{\alpha,\alpha}(\omega)= 
\left( \frac{ 4 e^2 t^2}{\omega} \right) e^{ -2 S_T} 
\left\langle \left( \frac{S_0+1/2}{2 S+1} \right) \right\rangle^2
\int_{- \infty}^{+ \infty} d \xi 
 [n_F(\xi-\omega)-n_F(\xi)]
\tilde{C}(\xi,\omega) h(\xi)
\label{130ar}
\end{equation} 
where $\tilde{C}(\xi,\omega)$ is    

\begin{equation}
\tilde{C}(\xi,\omega)=  
\frac{ \Gamma(\xi) }
{ \Gamma^2(\xi)+\omega^2 }
\label{131ar}
\end{equation} 
and $h(\xi)$ reads

\begin{equation}
h(\xi)=\left( \frac{1}{N} \right) \sum_{{\bf k}} \sin^2(k_{\alpha})
\delta( \xi - \xi_{{\bf k}} ).
\label{132ar}
\end{equation} 
The latter term of the conductivity becomes

\begin{eqnarray}
\Re \sigma^{(incoh)}_{\alpha,\alpha}(\omega)= && 
\left( \frac{ 2 e^2 t^2}{\omega} \right) e^{ -2 S_T} 
\left\langle \left( \frac{S_0+1/2}{2 S+1} \right)^2 \right\rangle
\int_{- \infty} ^{+ \infty} d \xi 
\int_{- \infty} ^{+ \infty} d \xi_1  
g(\xi) g(\xi_1) R(\xi,\xi_1,\omega)+
\nonumber \\
&& \left( \frac{ 2 e^2 t^2}{\omega} \right) e^{ -2 S_T} 
\left[ I_0(z) \left\langle \left( \frac{S_0+1/2}{2 S+1} \right)^2  
\right\rangle
- \left\langle \left( \frac{S_0+1/2}{2 S+1} \right)  \right\rangle^2  \right]
\times
\nonumber \\
&& \times
\int_{- \infty} ^{+ \infty} d \xi 
\int_{- \infty} ^{+ \infty} d \xi_1  
g(\xi) g(\xi_1)  [n_F(\xi-\omega)-n_F(\xi)]
C(\xi,\xi_1,\omega) 
\label{134ar}
\end{eqnarray}
where $g(\xi)$ is the density of states (\ref{12r}) and the function
$ R(\xi,\xi_1,\omega) $ is given by 

\begin{equation}
 R(\xi,\xi_1,\omega)=
2\sum_{l=1}^{+ \infty} I_l(z) 
\sinh \left( \frac{\beta \bar{\omega}_0 l}{2} \right)
\left[ 
J_l(\xi,\xi_1,\omega)+
H_l(\xi,\xi_1,\omega)
\right].
\label{135ar}
\end{equation}
We notice that $J_l( \xi,\xi_1,\omega )$

\begin{eqnarray}
J_l( \xi,\xi_1,\omega )=
C( \xi,\xi_1,\omega+l\bar{\omega}_0 ) 
[n_F(\xi-l\bar{\omega}_0-\omega)-
n_F(\xi-l\bar{\omega}_0)]
\left[ N_0(l \bar{\omega}_0)+n_F(\xi) \right]
\label{136ar}
\end{eqnarray}
and $H_l( \xi,\xi_1,\omega )$ 
 
\begin{eqnarray}
H_l( \xi,\xi_1,\omega )=
C( \xi,\xi_1,\omega-l\bar{\omega}_0 )
[n_F(\xi+l\bar{\omega}_0-\omega)-
n_F(\xi+l\bar{\omega}_0)]
\left[ N_0(l \bar{\omega}_0)+1-n_F(\xi) \right]
\label{137ar}
\end{eqnarray} 
describe phonon absorption and emission processes, respectively. 

In the limit of high temperatures ($T>0.39 \omega_0$) and for $SP$ 
excitations, the incoherent absorption is prevalent. In this case the 
conductivity consists of a sum of narrow Lorentzian functions
centred on the points $n \omega_0$ respectively. \cite{mahan} 
A derivation suitable for high temperatures can be performed and gives  
\cite{loosh} 

\begin{eqnarray}
&& \Re \sigma_{\alpha,\alpha}(\omega)= 
\nonumber \\
&&
\left( \frac{e^2 t^2}{\omega} \right)  
\left\langle \left( \frac{S_0+1/2}{2 S+1} \right)^2 \right\rangle
\sqrt{ \frac{ \pi \beta}{ \bar{z}}}
\left\{
\exp \left[  - \frac{\beta}{4 \bar{z}} 
( \omega - \bar{z} )^2  \right] -
\exp \left[  - \frac{\beta}{4 \bar{z}} 
(\omega + \bar{z} )^2  \right]
\right\}
\rho ( 1- \rho )
\label{60ar}
\end{eqnarray}
where $\rho$ is the electron concentration.
 
The conductivity $\sigma_{\alpha,\alpha}$ is given
by eq.(\ref{68r}) and eq.(\ref{60ar}) in $LP-FM$ and $SP-PI$ phase, 
respectively. In the regime of 
coexisting phases, the two preceding conductivities are combined by means of 
the eq.(\ref{14r}). 

At low temperatures, in the $FM$ phase the system shows two 
types of optical response: a Drude term and a broad absorption band due 
respectively to the coherent and incoherent motion of large polarons.
With increasing $T$, the optical spectra are characterized by a transfer of 
spectral weight from low to high energies.
At high $T$ the infrared absorption consists in a band peaked 
approximatively around the energy $2 g^2 \omega_0$ and is due to the incoherent
$SP$ dynamics. \cite{yoon,machida} 
The experimental data can be fitted reasonably well with the $LP$ and $SP$ 
spectra below $0.8$ $eV$. \cite{kim1,kim2,okim}
Indeed we note that for high frequencies the effects due to the exchange-split
bands and to the local Coulomb repulsion should be included. \cite{held}   
Thus the results can be considered meaningful for frequencies up to the 
absorption peak of the $SP$ band.

\section{Effects of an external magnetic field}
In this section we discuss the effects of an external magnetic field 
on the optical properties of the system at $x=0.3$.

We first note that a unit of magnetic field $h_0$ can be fixed as
\begin{equation}
h_0=\frac {\omega_0}  {  g_{s} \mu_{B} S}.    
\label{11r4}
\end{equation}
If we choose for  $\omega_0$ a reasonable value of $50 meV$, this unit is 
huge: in fact $h_0$ is of the order of $300$ $ T$. For realistic values of the 
magnetic fields the new energy scale $g_{s} \mu_{B} S h_0$  is small 
when compared with $\omega_0$.
However in presence of an external magnetic field the phase diagram shows 
the tendency for $LP-FM$ and $FM-PI$ regions to grow. 
By introducing an external magnetic field (see Fig. 2), the magnetization does
not vanish at $T_c$, the transition temperature in absence of the field, 
showing the behavior of a first order transition. \cite{zhou,shin}
Thus, in the vicinity of the transition temperature, the subtle balance 
between the two coexisting phases can be readily influenced by varying the 
magnetic field. 

The increase of the coherent motion due to the introduction of the external 
field is clearly shown in the calculated conductivity. We limit ourselves to 
the diagonal component. In Fig. 3(a) we can notice the rise of the Drude term 
with increasing the magnetic field, while in Fig. 3(b) the small-polaron
absorption band is suppressed. This behavior is more pronounced at higher 
temperatures (see Fig. 4 for $T=0.96 T_c$). At $T_c$ the 
introduction of the magnetic field is able to fill up the low frequency 
optical gap present in the high-temperature phase (Fig. 5).   
Hence the magnetic field induces the transfer of spectral weight from
high to low energies near the phase boundary. \cite{kaplan} 

The introduction of the magnetic field has dramatic consequences on the 
resistivity  
$\rho=1/\sigma_{\alpha,\alpha}$  near $T_c$, where 
$\sigma_{\alpha,\alpha}$ results from the conductivity 
$\sigma_{\alpha,\alpha}(\omega)$ in the limit $\omega \rightarrow 0$ (Fig. 6).
We obtain two different behaviors: metallic 
($d\rho/dT>0$) and insulating ($d\rho/dT<0$) emphasizing in logarithmic scale
the behavior of activated resistivity characteristic of a $SP$ phase
for $T>T_c$. \cite{schiffer,worledge}
The magnetic field shifts the peak of the resistivity at higher temperatures.
Since the $PS$ regime is characterized by a rapid increase of the resistivity,
\cite{perroni} 
at $T_c$ a large difference between resistivities at different magnetic fields
occurs.
The magnetoresistance ratio (see in Fig. 7(a) $MR$ in percent)
\begin{equation}
 MR(h)=\frac{ 
\left[ \rho(h=0)-\rho(h) \right] }
{\rho(h)}
\label{135ar1}
\end{equation} 
assumes large values at the transition temperature, so that the effect of 
the magnetic field is dramatically amplified near $T_c$. \cite{jin,schiffer}
Therefore, the $CMR$ effect and, in general, the high sensitivity of 
the system to external parameters are due to the subtle balance between 
$LP-FM$ and $SP-PI$ phases characterized by different lattice, spin 
and charge properties.

We have calculated the effect of the applied magnetic field on the $DW$ 
factor. In Fig. 7(b) we report $\Delta (DW)$, the relative variation of the 
$DW$ factors obtained without and with the external field, in order to make a 
comparison with the magnetoresistance ratio.  \cite{meneghini,meneghini1}   
The quantity $\Delta (DW)$ shows a well defined peak around the transition 
temperature, indicating that the application of the magnetic field reduces the
polaronic distortions. The two quantities, $MR$ and $\Delta (DW) $ in Fig. 7, 
 are both peaked around $T_c$ showing the existence of a strong correlation 
between lattice distortions, conductivity and magnetic properties. 
\cite{booth2}

\section{Isotope effect}
In this section we deal with the isotope effect in the $CMR$ regime at $x=0.3$.
In particular we calculate the effects of isotope substitution on the infrared
spectra.  

The isotopic substitution of the oxygen from $^{16} O$ to $^{18} O$ changes 
the values of $\omega_0$ and $g$ to $\omega_0^*=\omega_0\sqrt{M/M^*}$ and 
$g^*=g\left(M^*/M\right)^{1/4}$ respectively, where $M^*$ is the mass of 
$^{18} O$. The coupling parameter $\lambda= g^2 \omega_0 / 6t $ is clearly 
unchanged.   

In Fig. 8 the phase diagram for $^{18} O$ is presented and compared with that
of  $^{16} O$. It is qualitatively 
altered with respect to the phase diagram of $^{16} O$  showing a reduction 
of the $FM$ regions. \cite{5,isaac,zhao1,zhao2,ibarra,kresin,lorenz}
The heavier ion mass of $^{18} O$ reduces the effective transfer integral 
$t_{eff}$ (see eq. (\ref{10r})), so the region of coexisting phases and in 
particular the Curie temperature $T_c$ decrease. Indeed, in the limit where 
the intra-atomic exchange integral $J$ is far larger than the bare transfer 
integral $t$, $T_c \propto t_{eff}$.\cite{3,zhang1} In the inset of Fig. 8, 
$T_c^*$, the Curie temperature for $^{18} O$, is reported along with $T_c$, 
the Curie temperature for $^{16} O$.
We have evaluated the oxygen-isotope exponent $\alpha_0=-\Delta ln T_c/ 
\Delta ln M=- \frac{(T_c^*-T_c)}{T_c} \frac{M}{(M^*-M)}$, making a comparison 
with experimental values. \cite{isaac}  
We find the decreasing behavior of $\alpha_0$ with increasing the hole doping 
$x$ (Fig. 9).

Near $T_c$ we have calculated the optical conductivity for
the two isotopes (Fig.10). While the $^{16} O$ system is in the $PS$ regime, 
$^{18} O$ system is in the insulating phase. In the first case the optical 
response still shows the Drude term, for $^{18} O$ only the small-polaron band
at high frequency is present. Again this causes large changes in the 
conductivity of the system. Therefore the isotope effect represents another 
example of the sensitivity to the variation in external parameters.

\section{Conclusions} 

We have discussed the effects of an external magnetic field and of the oxygen
isotopic substitution on the infrared absorption spectra  for $x=0.3$ 
mainly near the phase boundary. Furthermore we have dealt with the 
lattice effects and the $DW$ factor without and with the magnetic field.

First we have focused our attention on $LDPDF$ finding that, near and in the 
$PI$ phase, only the sites occupied by an electron show strong lattice 
distortions and that the $DW$ factor increases rapidly with 
temperature near $T_c$ in agreement with $XAFS$ measurements. 
\cite{tyson,booth1,booth,booth2}

Then we have examined the consequences of an external magnetic field on the 
infrared absorption spectra at $x=0.3$ stressing that, 
near the phase boundary, the application of the field induces a transfer of 
spectral weight from high to low energies \cite{kaplan} and evaluating the 
$CMR$ ratio. 
In our scheme, the found $CMR$ is due to the subtle balance between  
coexisting phases characterized by different lattice, spin and electronic 
properties.
The relation between the $CMR$ ratio and the variation of the $DW$ 
factor in presence of the magnetic field has pointed out the correlation that
involves lattice distortions, conductivity and magnetic properties.

Finally, within our approach, the oxygen isotope effect has been 
explained by the enhanced electron-phonon coupling that induces a reduction of
the $FM$ phases. In the vicinity of the $MI$ transition the isotope 
substitution induces large changes in the optical response and in the 
resistivity in agreement with experimental data. \cite{5,isaac,zhao1,zhao2}
 
We believe that the strong sensibility of the infrared absorption with
respect to the magnetic field and isotopic substitution could be used to test 
the validity of the $PS$ scenario adopted in this paper. Indeed the 
experimental study of the infrared absorption is well established and it could
be able to observe the effects discussed above. 

In our work, near the $MI$ transition, the interplay of the electron-phonon 
interaction and the magnetic effects gives rise to a charge-density 
instability that takes into account experimental evidences for coexisting 
localized insulating and delocalized metallic components. 
\cite{9a,9b,9c,fath,adams,zuo,jaime}  
We have assumed that the scales of the inhomogeneities are much 
larger than the inter-particle distance and that the term arising from the 
mixing energy of the two coexisting phases and including the surface energy 
cost is not able to change qualitatively the behavior of the macroscopic
quantities discussed. 
We have exploited a macroscopic $PS$ scenario where the coexisting fractions 
of volume are determined by the variational procedure and the Maxwell 
construction. 
This approach to manganites provides results consistent with experimental 
measurements \cite{cata,perroni} and an explanation of the strong sensitivity
of the system to external parameters.

\section*{Figure captions}
\begin{description}

\item  {F1} 
(a) The distribution function of the atomic displacement 
(in units of $\sqrt{ M \omega_{0}/ 2}$)  at $x=0.3$ as a function of the 
displacement coordinate $X$ (in units of $2/ \sqrt{ M \omega_{0}}$ different
from the length unit introduced above, the lattice constant $a$).

(b) The Debye-Waller factor (in units of $2/ M \omega_0)$  
  at $x=0.3$ as a function of the temperature. 

\item  {F2} 
The magnetization (in units of saturation magnetization $M_{S}$) as a function
of the temperature for two different magnetic fields (in units of $h_0$).

\item  {F3} 
(a)-(b) The diagonal conductivity  at $T=0.9$ $ T_{c}$ at different ranges of 
the frequency $\omega$  for several magnetic fields (in units of $h_0$).
The conductivities are expressed in units of  $ e^2 c/ m \omega_0 $,
with $c$ hole concentration and $m= 1/ 2t $.

\item  {F4} 
(a)-(b) The diagonal conductivity (in units of  $ e^2 c /m \omega_0$,
with $c$ hole concentration and $m= 1 / 2t$) at $T=0.96$ $ T_{c}$ 
at different ranges of the frequency $\omega$ for several 
magnetic fields (in units of $h_0$).

\item  {F5}
The diagonal conductivity up to 18 $\omega_0$ at $T=1.02$ $ T_{c}$ for 
different magnetic fields (in units of $h_0$).
The conductivity is expressed in units of $ e^2 c / m \omega_0$,
with $c$ hole concentration and $m= 1 / 2t $.

\item  {F6} 
The inverse of the diagonal conductivity as a function of the temperature at 
different magnetic fields (in units of $h_0$) (we have used 
$\omega_0=50$ $meV$ and the lattice constant $a=0.4$ $nm$).

\item  {F7} 
(a) The magnetoresistance ratio at a fixed magnetic field 
(in units of $h_0$) as a function of the temperature.

(b) The relative variation of the Debye-Waller factor at a fixed magnetic 
field (in units of $h_0$) as a function of the temperature.

\item  {F8} 
The phase diagram for the two different oxygen isotopes
($ ^{16}O $ dashed line, $ ^{18}O $ dotted line ). $PI$ means Paramagnetic 
Insulator, $FM$ Ferromagnetic Metal and $AFI$ AntiFerromagnetic Insulator.
The areas $PI+FM$ and $AFI+FM$ indicate regions where localized ($PI$ or 
$AFI$) and delocalized ($FM$) phases coexist.
In the inset the ferromagnetic transition temperatures as a function of the 
hole doping ($ T_{c}$ dashed line for $ ^{16}O $ and $ T_{c}^{*}$ dotted line 
for $ ^{18}O $). The temperatures are expressed in units of $\omega_0$ and
the model parameters are $t=2 \omega_0$, $g = 2.8$ ($\lambda = 0.65 $ for 
both isotopes) and $\epsilon= 0.01 t$.

\item  {F9}
The oxygen-isotope exponent $\alpha_0$ (circles) compared to the experimental
exponent (diamonds) deduced by Ref. 33 as a function of the hole doping.

\item  {F10}
The conductivity  at $T=0.97$ $T_{c}$ up to 19 $\omega_0 $ for $^{16} O$ and
 $^{18} O$. The conductivities are expressed in units of  
$ e^2 c/ m \omega_0 $, with $c$ hole concentration and $m= 1/ 2t $.

\end{description}

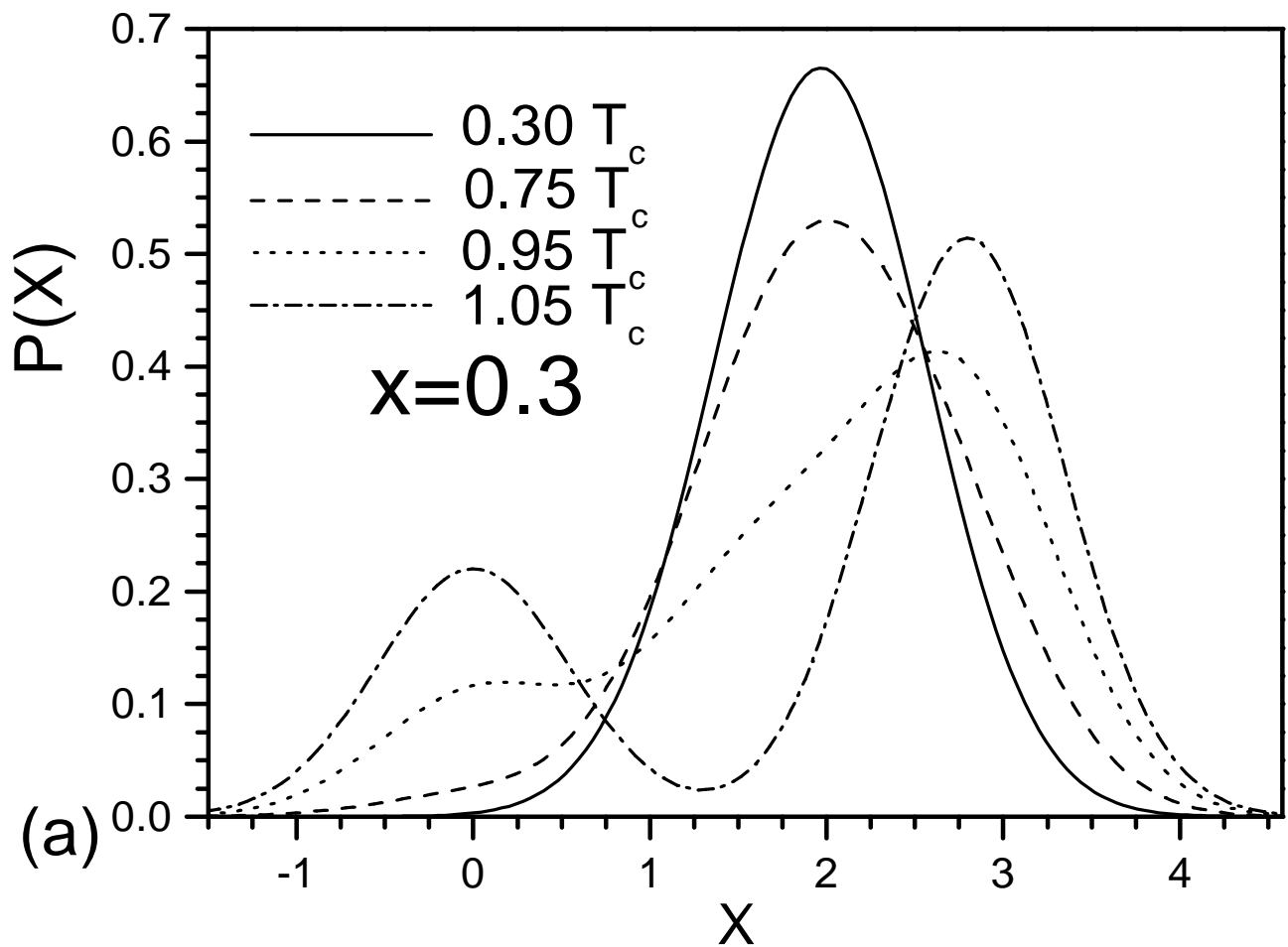
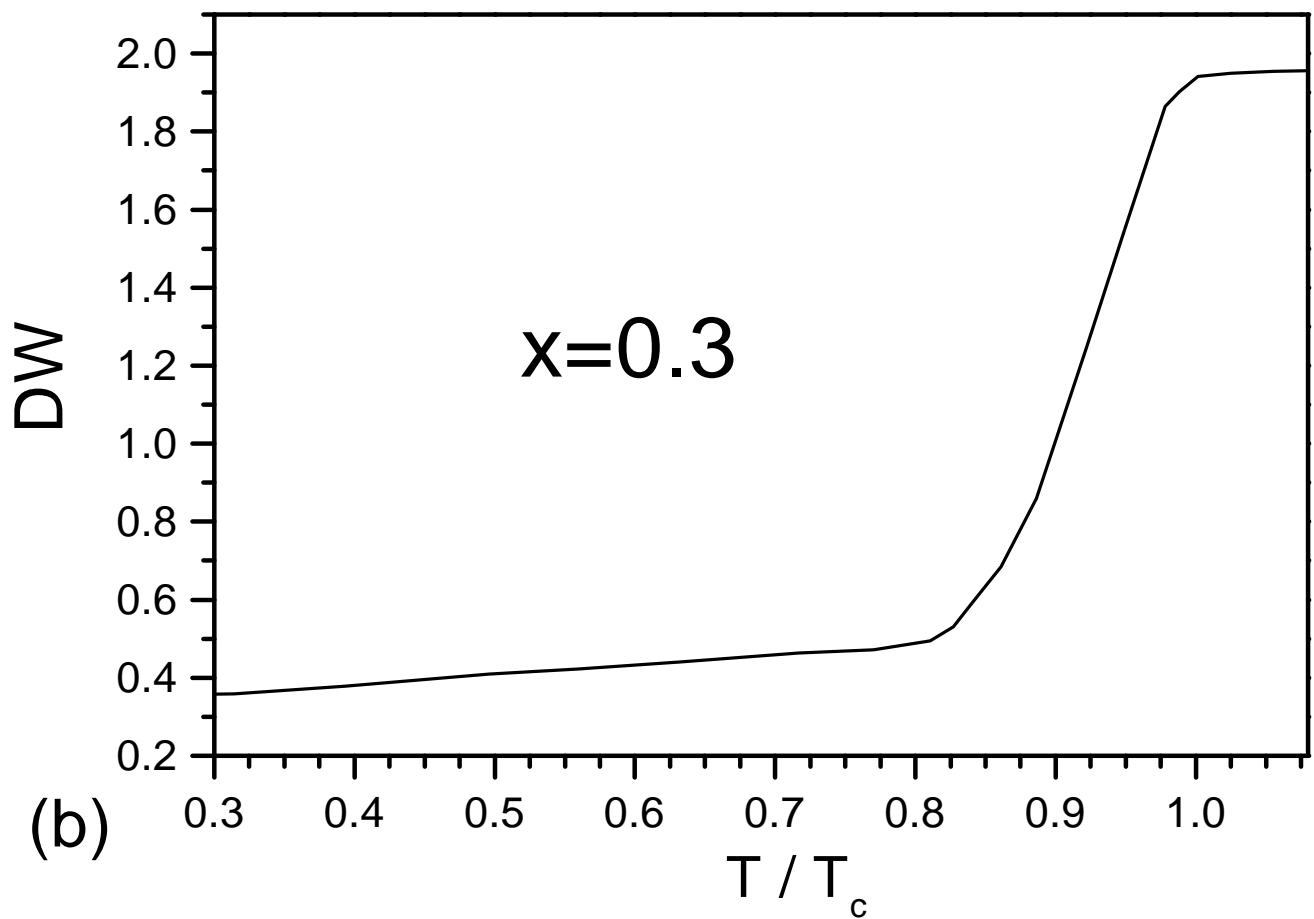
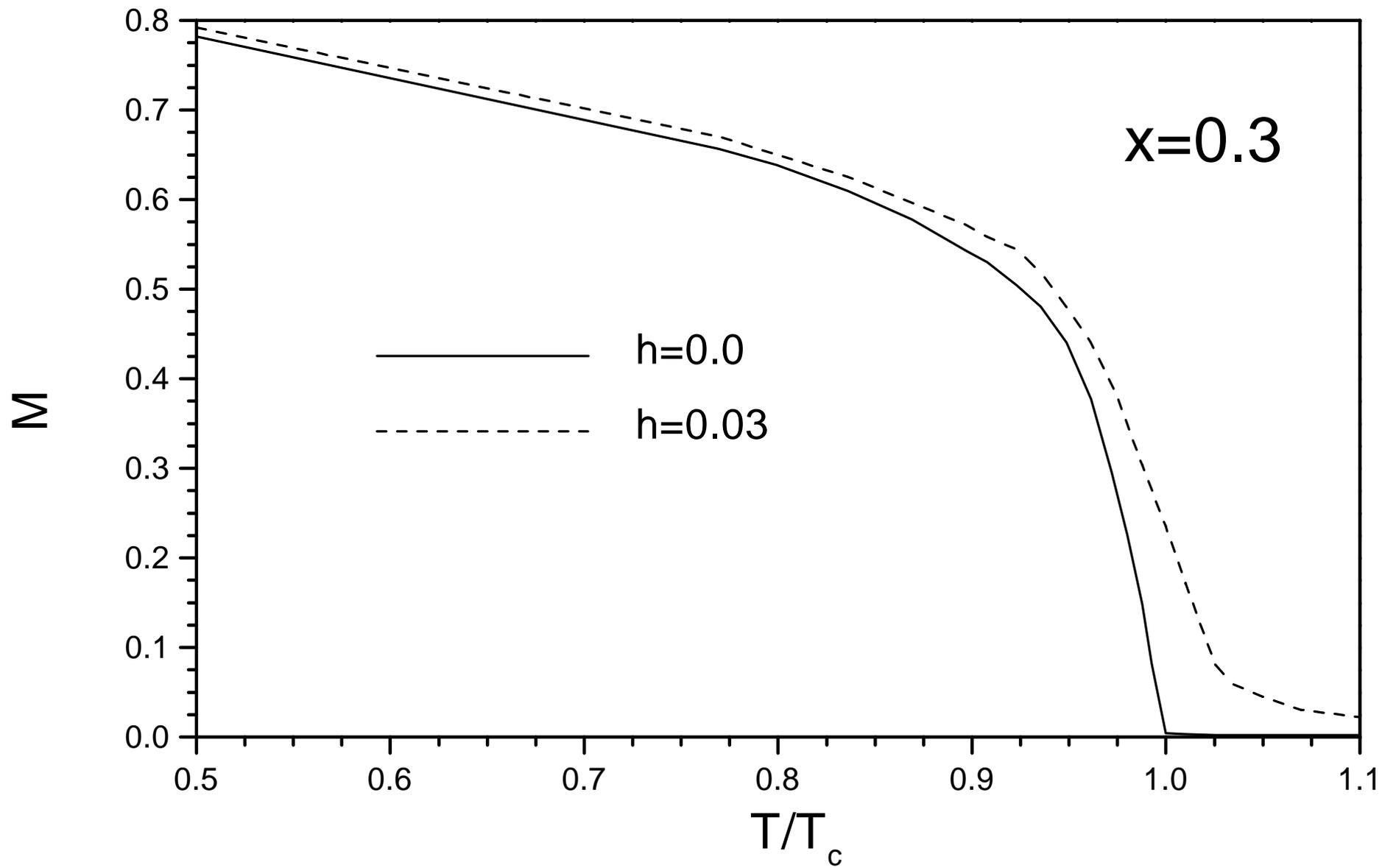
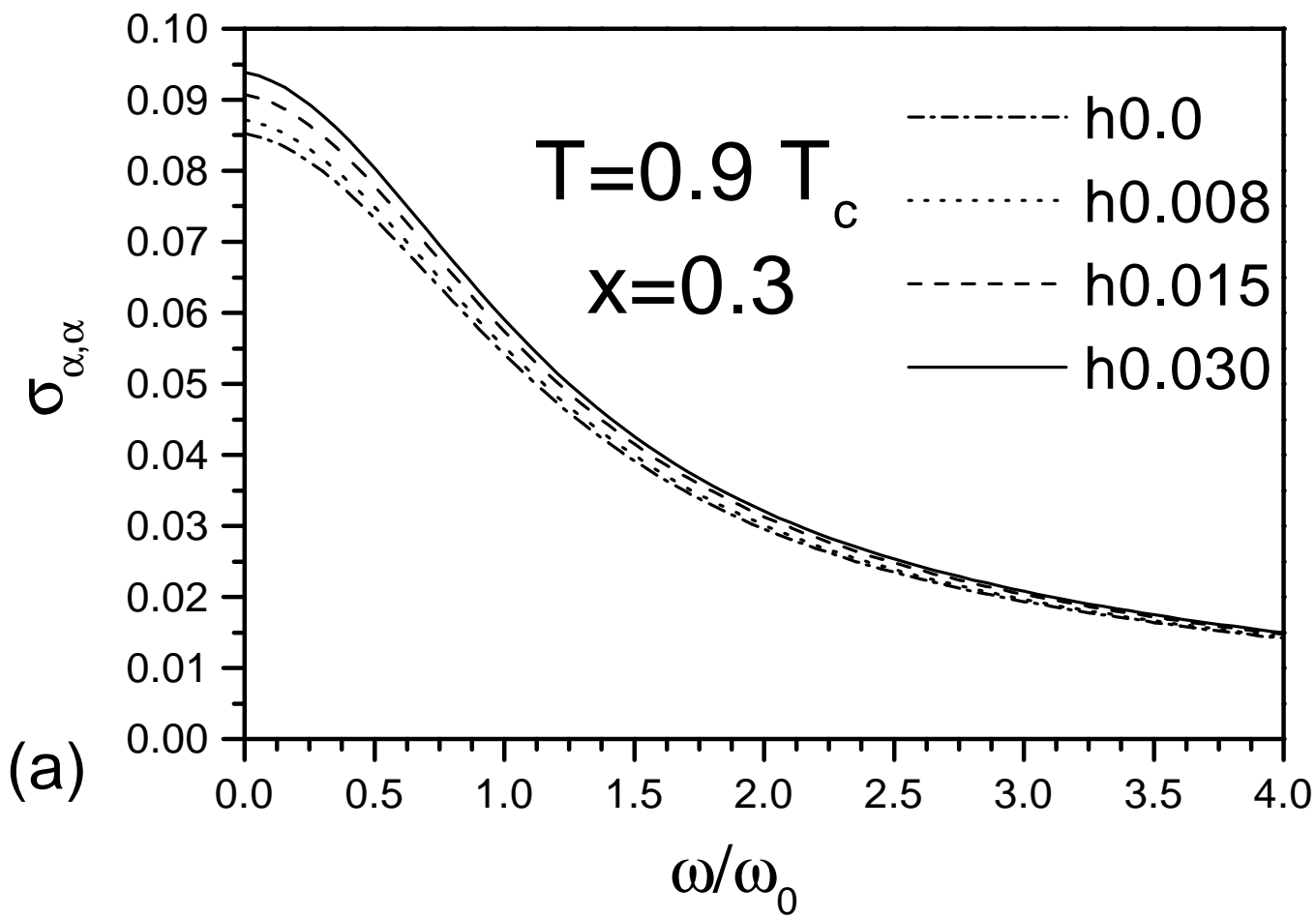
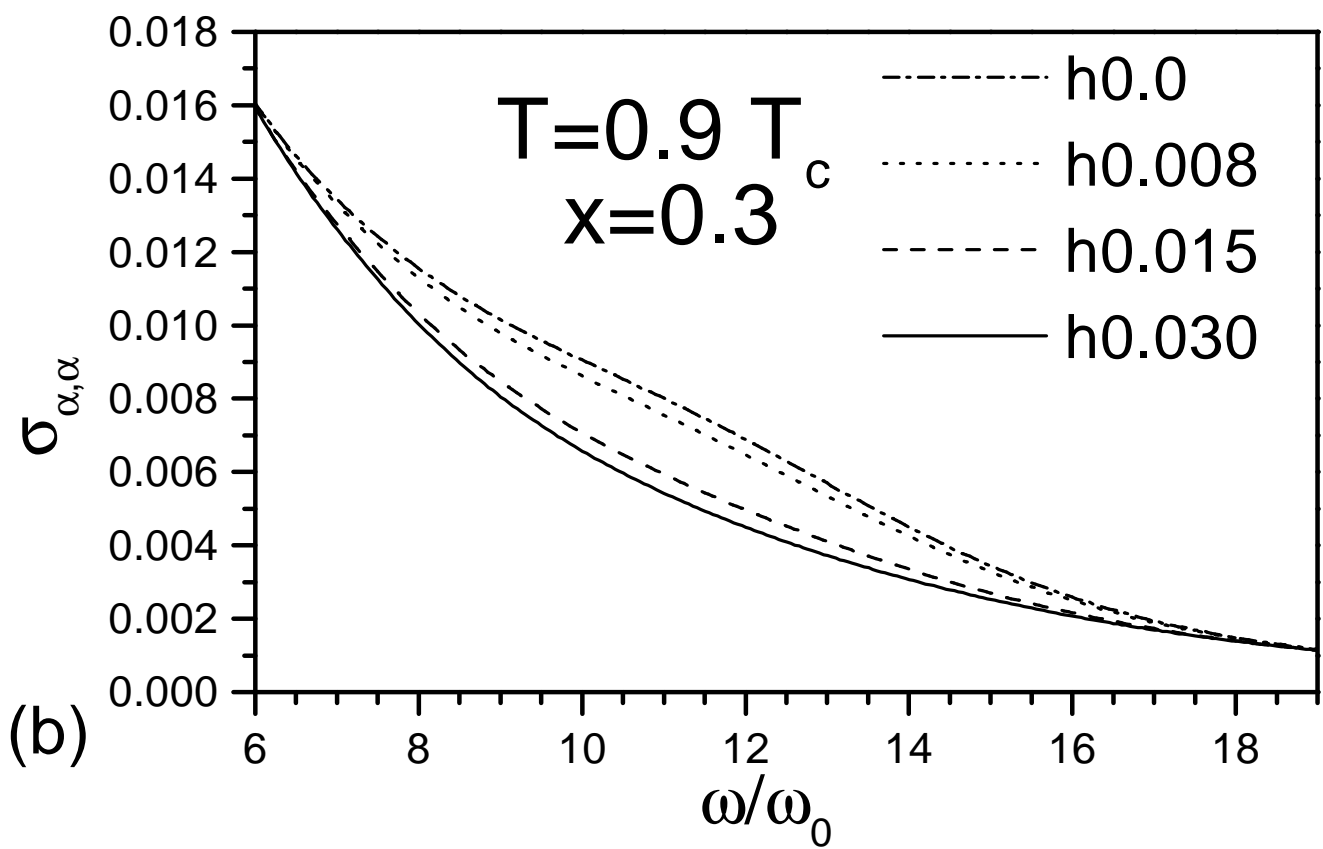
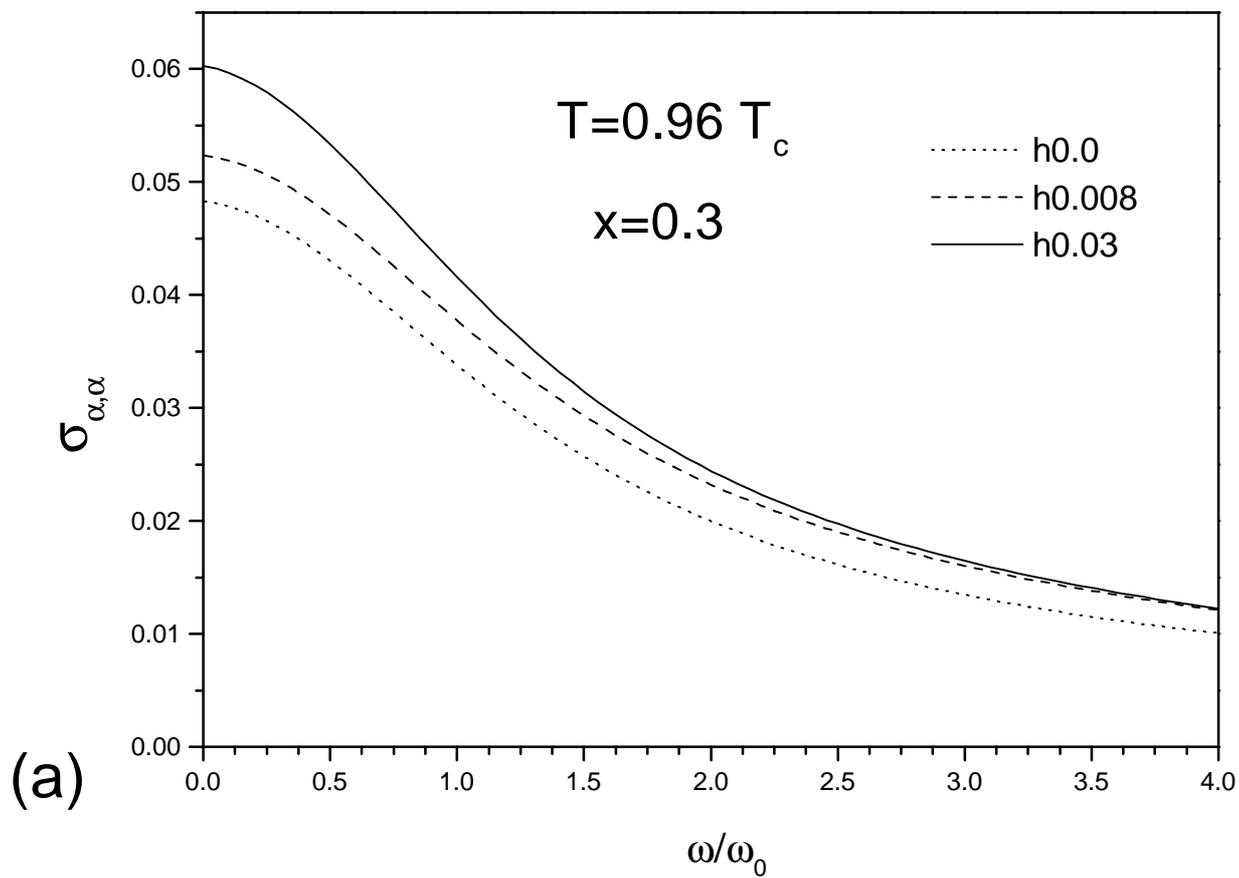
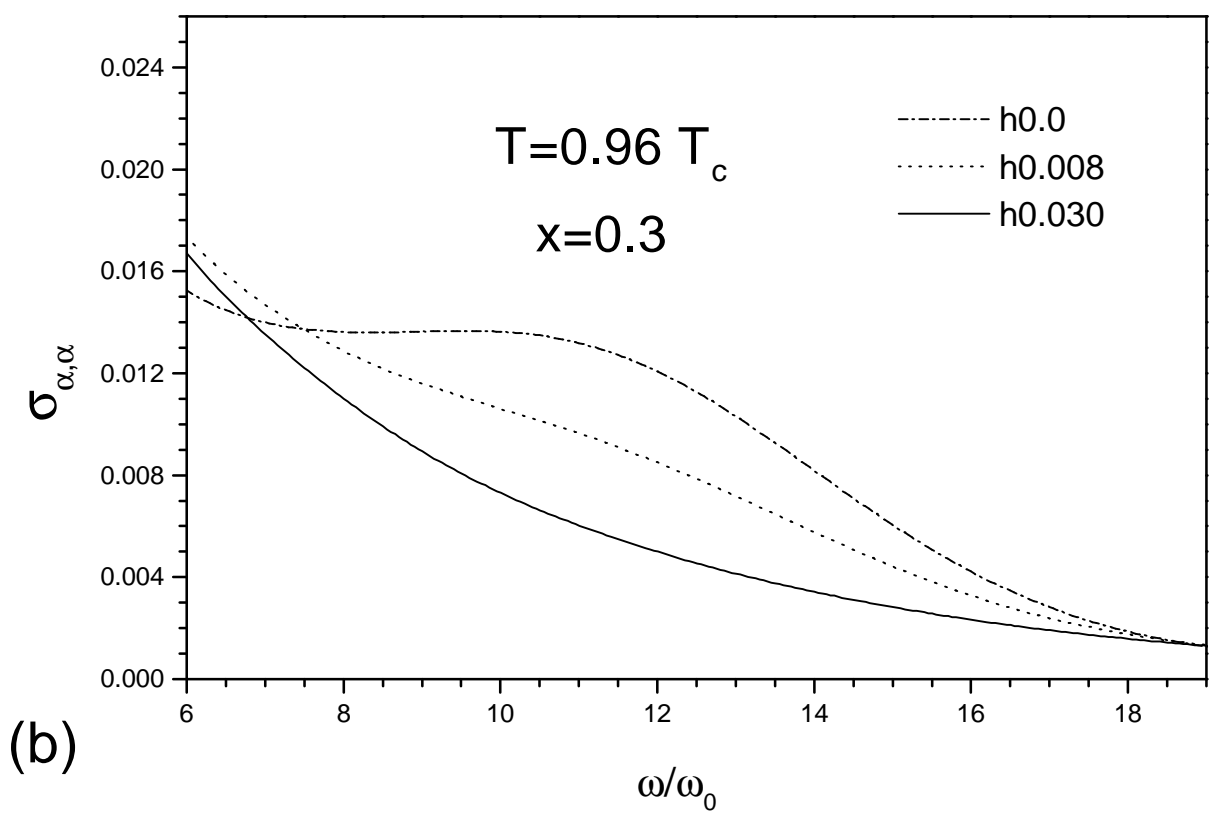
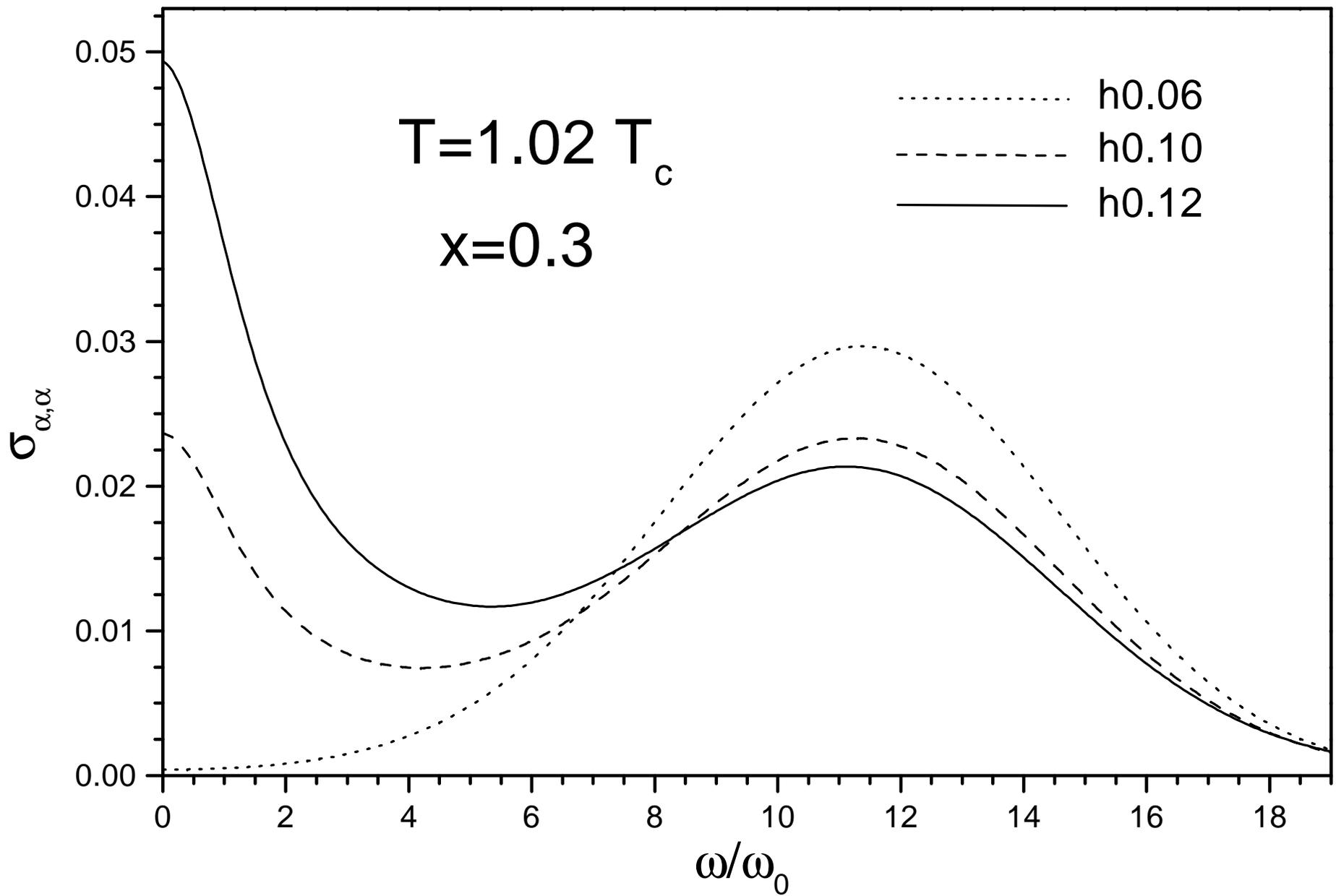
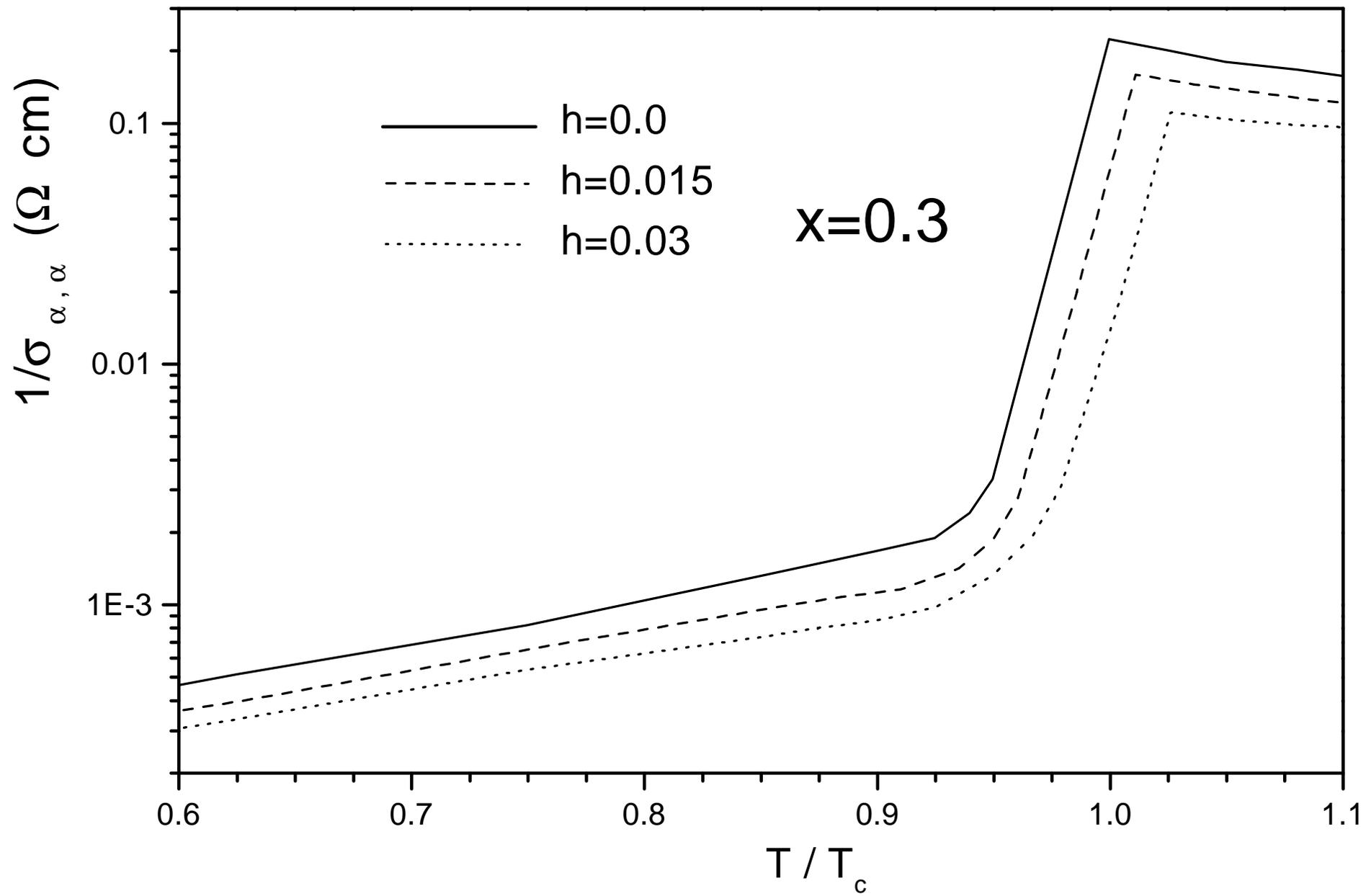
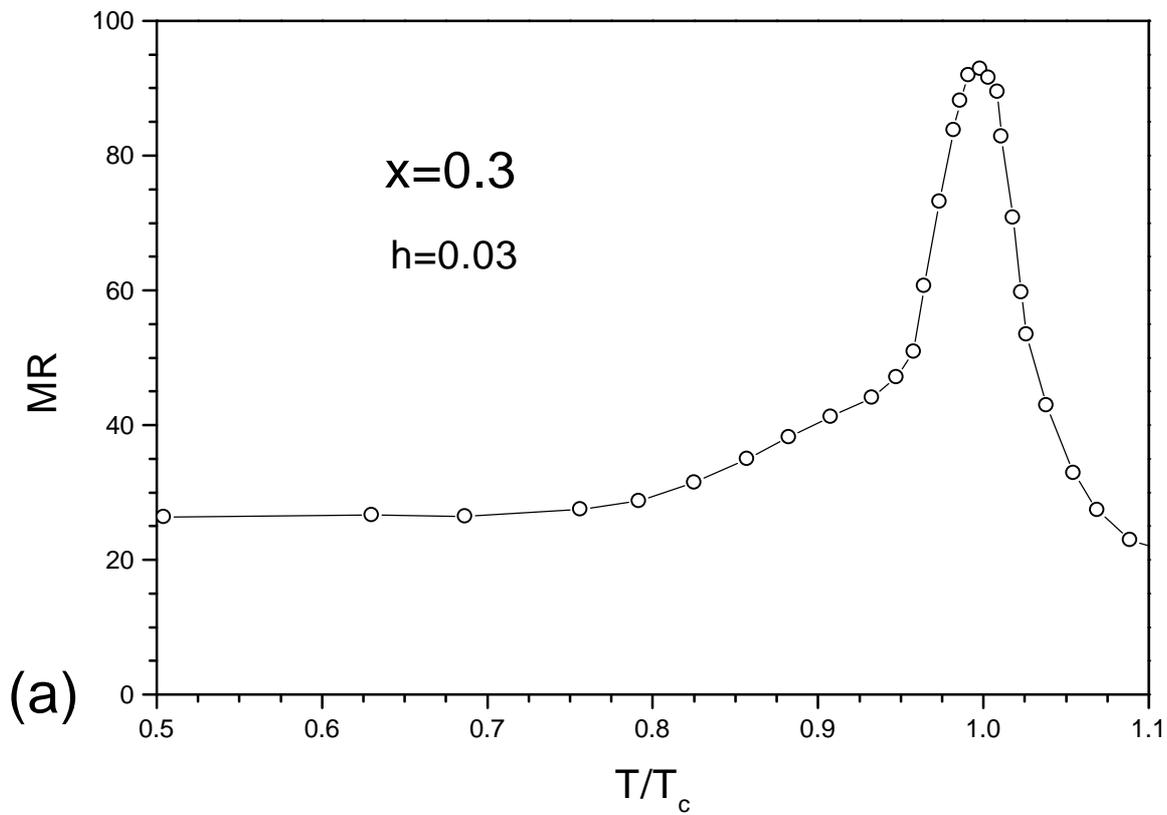
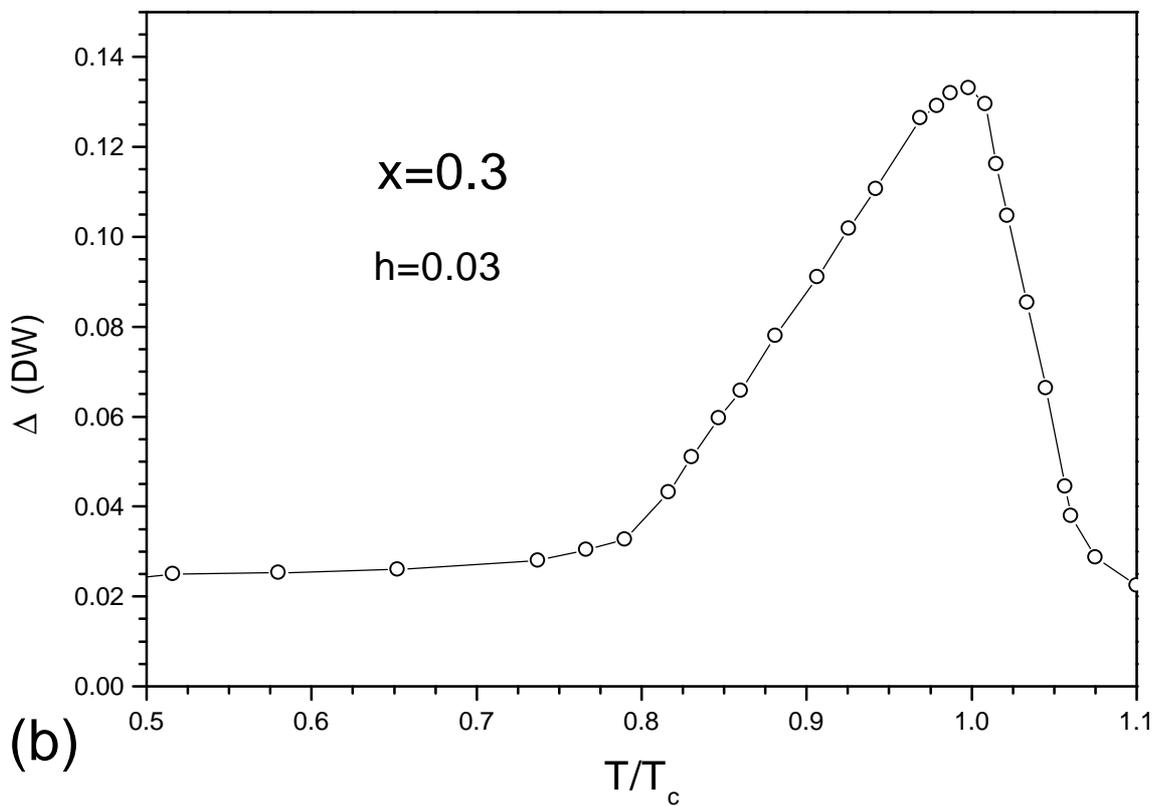
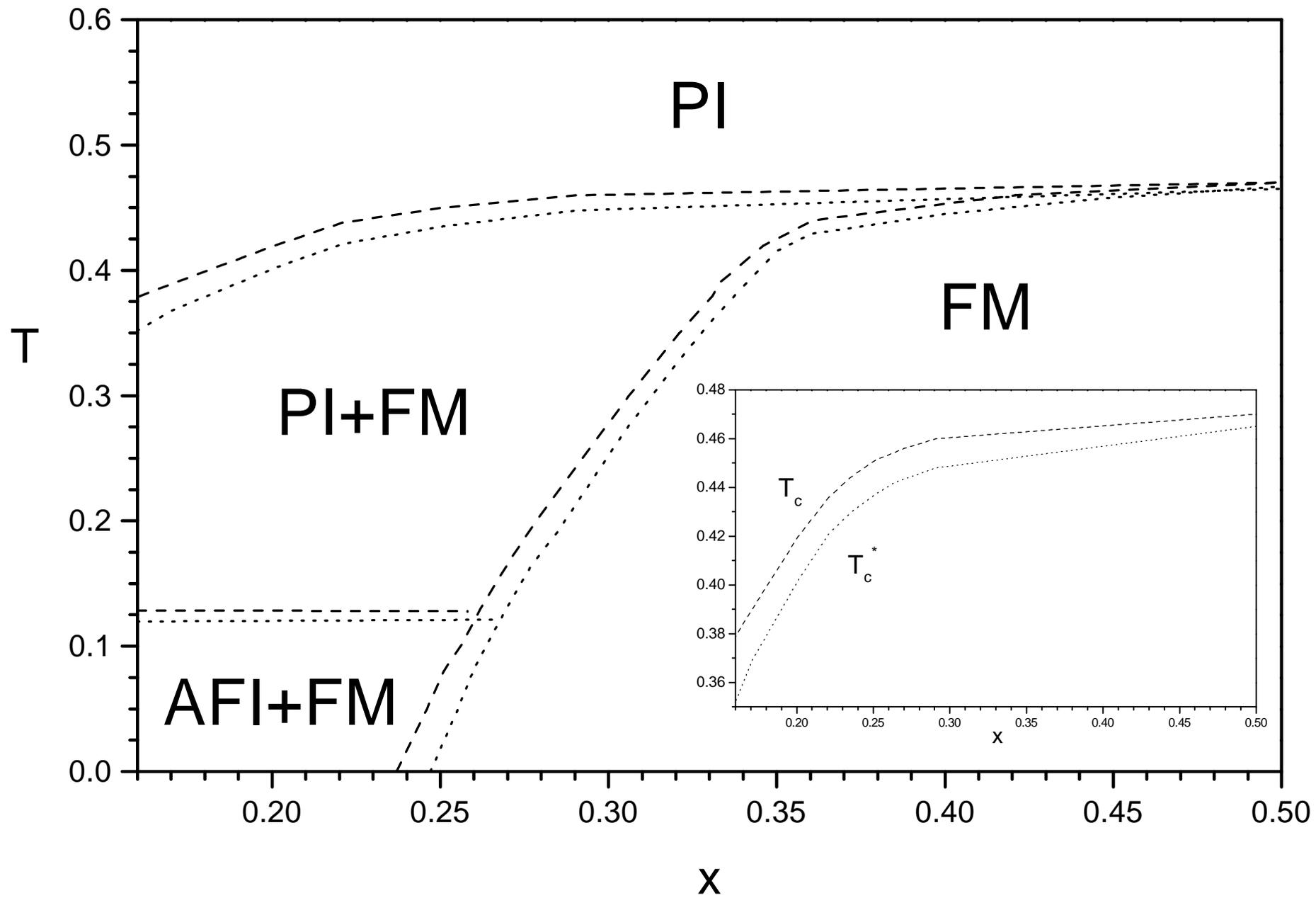
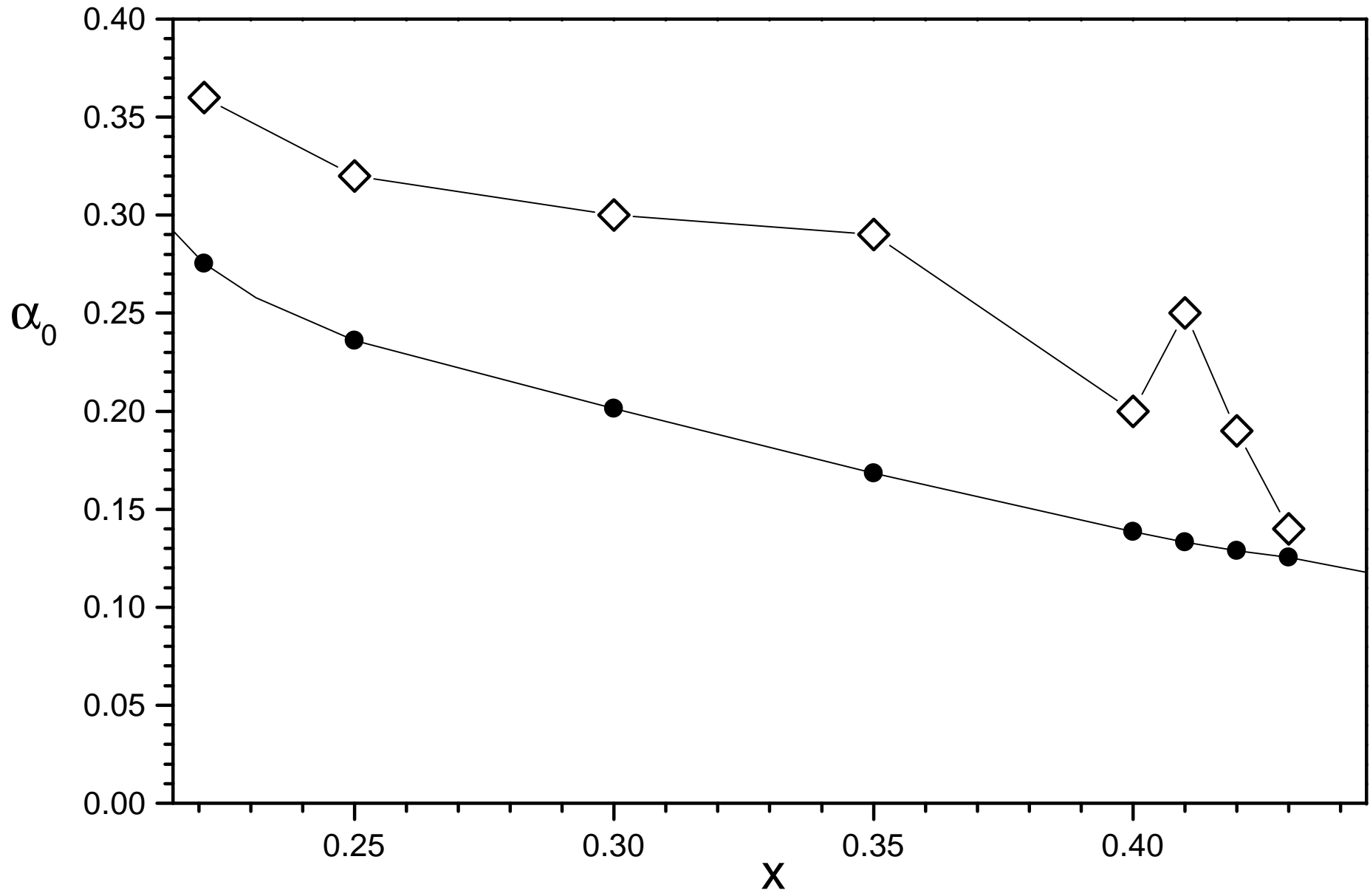
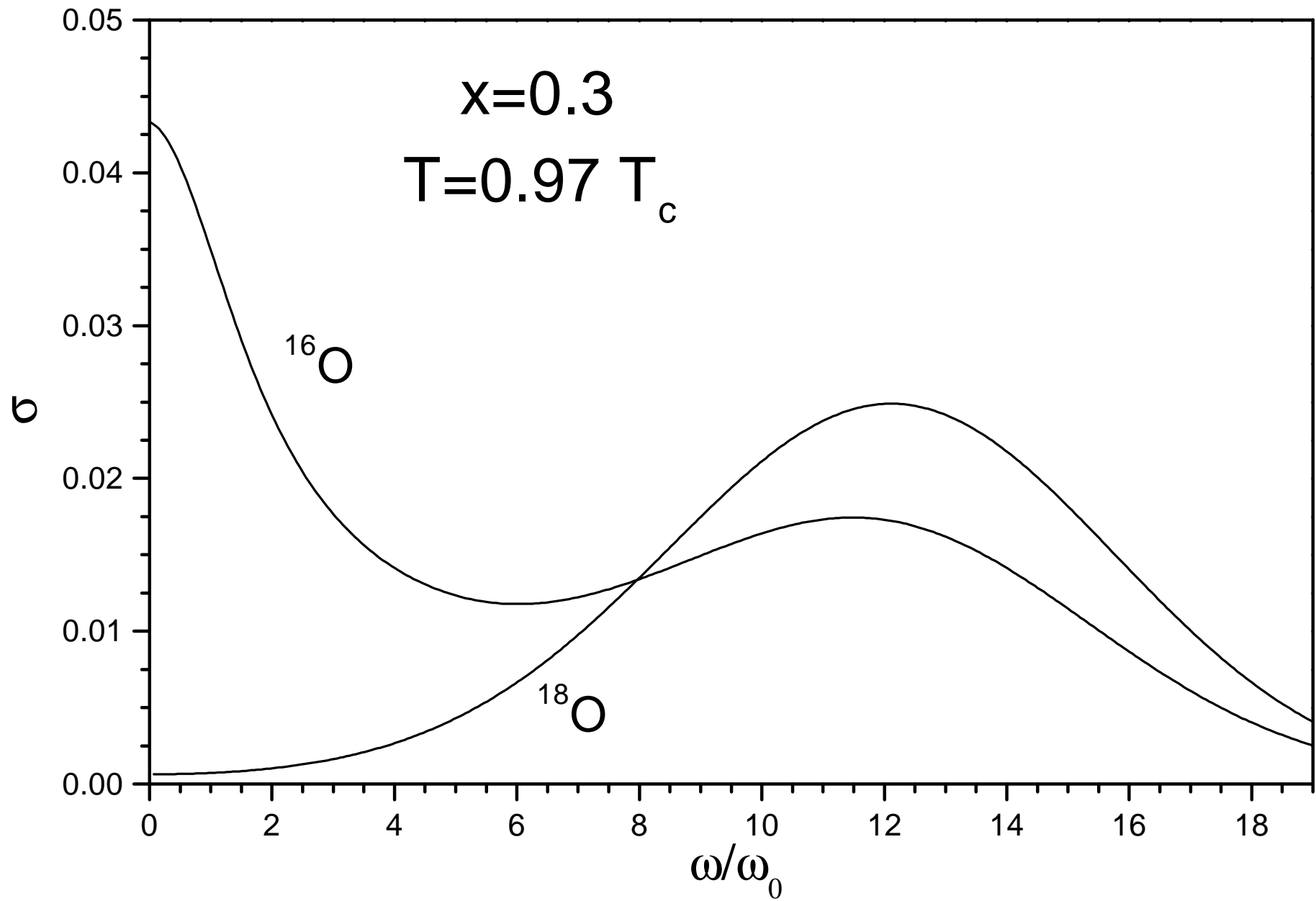
\end{document}